\documentclass{article}

\usepackage{amssymb,amsfonts,amsmath,stmaryrd}
\usepackage{cite,enumerate,float,indentfirst}
\usepackage{color}

\def\be{\begin{eqnarray}}
\def\ee{\end{eqnarray}}
\def\nn{\nonumber}

\def\p{\partial}
\def\tr{{\rm tr}\,}
\def\Tr{{\rm Tr}\,}

\newcommand{\beq}{\begin{equation}}
\newcommand{\eeq}{\end{equation}}
\newcommand{\beqa}{\begin{eqnarray}}
\newcommand{\eeqa}{\end{eqnarray}}

\definecolor{red}{rgb}{1,0,0}
\definecolor{orange}{rgb}{1,0.5,0}
\definecolor{violet}{rgb}{0.7,0,1}

\voffset= - 1.25in
\hoffset= - 1.0in         

\begin{document}

\title{\vspace{1.5cm}\bf
CMM formula as superintegrability property\\ of unitary model
}

\author{
A. Mironov$^{b,c,d,}$\footnote{mironov@lpi.ru,mironov@itep.ru},
A. Morozov$^{a,c,d,}$\footnote{morozov@itep.ru},
A. Popolitov$^{a,c,d,}$\footnote{popolit@gmail.com}
}

\date{ }

\maketitle

\vspace{-6cm}

\begin{center}
  \hfill MIPT/TH-22/24\\
  \hfill FIAN/TD-11/24\\
  \hfill ITEP/TH-28/24\\
  \hfill IITP/TH-23/24
\end{center}

\vspace{4.5cm}

\begin{center}
$^a$ {\small {\it MIPT, Dolgoprudny, 141701, Russia}}\\
$^b$ {\small {\it Lebedev Physics Institute, Moscow 119991, Russia}}\\
$^c$ {\small {\it NRC ``Kurchatov Institute", 123182, Moscow, Russia}}\\
$^d$ {\small {\it Institute for Information Transmission Problems, Moscow 127994, Russia}}
\end{center}

\vspace{.1cm}

\begin{abstract}
A typical example of superintegrability is provided by expression of the
Hopf link hyperpolynomial in an arbitrary representation through a pair of the Macdonald polynomials
at special points.
In the simpler case of the Hopf link HOMFLY-PT polynomial and a pair of the Schur functions, it is a relation in the unitary matrix model.
We explain that the  Cherednik-Mehta-Macdonald (CMM) identity for
bilinear Macdonald residues with an elliptic weight function is nothing but
a reformulation of these same formulas.
Their lifting to arbitrary knots and links, even torus ones remains obscure.
\end{abstract}

\bigskip

\newcommand\smallpar[1]{
  \noindent $\bullet$ \textbf{#1}
}

\section{Introduction}

Superintegrability is a remarkable property of integrable physical systems,
which expresses their properties through explicit single-valued functions and is related to additional integrals of motion, that typically have non-linear Poisson brackets
among themselves (hidden symmetries).
The best known classical example is the motion in Newton/Coulomb potential
when the orbits are closed and described in terms of simple elliptic integrals.
In other potentials, a precession of perihelion and other phenomena occur
which do not necessarily violate integrability of motion in the central field,
still make solutions of equations far more sophisticated.

Phenomenon of superintegrability
manifests itself in various problems.
In particular, it is quite interesting in the setting of (eigenvalue) matrix models,
which are always integrable \cite{GMMMO,KMMOZ,UFN3}, but not always superintegrable.
A typical example of superintegrability in this context is the identity
for the Schur function averages in Hermitian matrix model:
\be
\int_{N\times N} dX e^{\tr \sum_k \frac{p_k}{k}  X^k} S_\lambda(X) \ \stackrel{p_k=\delta_{k,2}}{=}\  \eta_\lambda(N) S_\lambda\{p_k\}\ ,
\label{SIherm}
\ee
where $X$ is the $N\times N$ Hermitian matrix, $S_\lambda(X)$ is the Schur function, which is a symmetric function of eigenvalues of the matrix $X$ labeled by the partition $\lambda$, and
$\eta_\lambda(N) = \frac{ S_\lambda(1)}{S_\lambda\{\delta_{k,1}\}}$
is a factorized function of $N$. A similar identity, when the average is factorized
is true for monomial potentials with $p_k=\delta_{k,r}$, for logarithmic potentials (Selberg integrals),
and in a few more sophisticated cases \cite{MMrev} but not in general.

Another kind of superintegrability in matrix models, which is called strict superintegrability \cite{MMsi,MMNek,CMPT} is a similar factorization property of bilinear averages, which, for the Hermitian matrix model case, takes the form
\be
\int_{N\times N} dX e^{\tr \sum_k \frac{p_k}{k}  X^k} S_\lambda(X)K_\mu(X) \ \stackrel{p_k=\delta_{k,r}}{=}\  \zeta_\lambda(N) S_{\nu(\lambda,\mu)}\{p_k\}\ ,
\label{SIherm2}
\ee
where $K_\nu(X)$ is an explicitly known basis in the space of symmetric functions (linear combinations of the Schur functions), $\zeta_\lambda(N)$ is another factorized function of $N$, and $\nu(\lambda,\mu)$ is a partition made of $\lambda$ and $\mu$ partitions.

For unitary matrix model, the simplest example of both superintegrability and strict superintegrability is given by the trivial potential and is contained in the orthogonality relation \cite{book:M-symmetric-functions-and-hall-polynomials}:
\be
\frac{1}{N!} \oint \prod_{i=1}^N {dz_i\over z_i}\prod_{i\ne j}^N \Big(1-{z_i\over z_j} \Big)
  \ S_\lambda(z)S_\mu(z^{-1})\ .
  =
  \delta_{\lambda,\mu}
\label{orthoU}
\ee
At the same time, non-trivial potentials typically give rise to non-factorized though controlled averages \cite{MMZ}.

Here we look at another example of unitary matrix model superintegrability, that with a non-trivial potential, it is associated with the expression for colored HOMFLY-PT polynomial of the Hopf link with two components colored with representations associated with the Young diagrams $\lambda$ and $\mu$\cite{AMMM}:
\be
\int_{N\times N}  [dU] e^{-\frac{1}{4\hbar} \tr (\log U)^2} S_\lambda[U] S_\mu[U] =
C_{\lambda,\mu} S_\lambda(q^{\mu_i-i+1})S_\mu(q^{1-i})
\label{uniHopf}
\ee
with $q:=e^\hbar$ and $C_{\lambda,\mu}$ constant,
which is actually symmetric under a permutation of $\lambda$ and $\mu$, and can be lifted to hyperpolynomials
expressed through Macdonald averages \cite{AKMMM,AKMM1}.
From knot/Chern-Simons theory point-of-view, it is conceivable that the l.h.s. is actually divisible by dimension of quantum representation
$D_{\lambda\cup\mu}(q^N,q)=S_{\lambda\cup\mu}(q^{i-1})$, what means that the factor $S_\lambda(q^{\mu_i+i-1})$ at the r.h.s.
is often factorized further.
This is an interesting fact, actually applicable to arbitrary knots, and it will be discussed elsewhere.
In the present paper, we concentrate on the generalizations of another kind.

The Gaussian potential in (\ref{uniHopf}) can be substituted by a theta-function  weight (see below),
with the help of Poisson transform
\be\label{Gth}
\oint\frac{dz}{z} \theta(z)\, z^m := \oint\frac{dz}{z} \left(\sum_k q^{k^2/2} z^k\right)z^m = q^{m^2/2}
\sim \int d\xi e^{-\frac{\xi^2}{2}} e^{im\xi} = \oint \frac{dz}{z} e^{-\frac{1}{2\hbar}(\log z)^2} z^m\ ,
\ee
where
\be
\theta(z):=\sum_{n=-\infty}^\infty q^{n^2/2}z^{n}\ .
\label{thetaf0}
\ee
In more detail, there are two formulas similar to (\ref{uniHopf}):
\be\label{eq:two-schurs-symmetric0}
\boxed{
\begin{array}{c}
  \frac{1}{N!} \oint \prod_{i=1}^N {dz_i\over z_i}\prod_{i\ne j}^N \Big(1-{z_i\over z_j} \Big)
  \ S_\lambda(z)S_\mu(z)
  \prod_{i=1}^N\theta(z_i)
  =q^{c^+_{\lambda,\mu}}
    \cdot S_\lambda(q^{\mu_i-i+1}) S_\mu(q^{1-i})\cdot\prod_{i>j}^N(1-q^{i-j})\ ,\cr
    \cr
      \frac{1}{N!} \oint \prod_{i=1}^N {dz_i\over z_i}\prod_{i\ne j}^N \Big(1-{z_i\over z_j} \Big)
  \ S_\lambda(z)S_\mu(z^{-1})
  \prod_{i=1}^N\theta(z_i)
  = q^{c^-_{\lambda,\mu}}
    \cdot S_\lambda(q^{-\mu_i+i-1}) S_\mu(q^{i-1})\cdot\prod_{i>j}^N(1-q^{i-j})\ ,
\end{array}
}
\ee
and $c^-_{\lambda,\mu} = \frac{1}{2} (|\lambda| + |\mu|)+\kappa_2(\lambda)+\kappa_2(\mu)$,
 $c^-_{\lambda,\mu} ={N(N-1)^2\over 6}+(N-1)(|\lambda| + |\mu|)+c^-_{\lambda,\mu}$
are expressed through the eigenvalues of the first  and the second Casimir operators:
$\kappa_1(\lambda)=|\lambda|$ and
$\kappa_2(\lambda)=\sum_{i,j\in \lambda}(j-i)=
\frac{1}{2}\sum_i \left(\left(\lambda_i-i+\frac{1}{2} \right)^2- \left(-i+\frac{1}{2}\right)^2\right)$.
The second of these identities can be considered as a profound deformation with an elliptic (theta-function) weight
of orthogonality relation for the Schur functions (\ref{SIherm2}). The r.h.s. of these two averages are the HOMFLY-PT polynomials for the Hopf links related to each other by reflection.

Remarkably, (\ref{eq:two-schurs-symmetric0}) is a much more direct counterpart of (\ref{SIherm}) and (\ref{SIherm2})
than it can seem, because it gives rise to the identities
\be\label{nb}
\int dU\exp\left(\sum_{k\in\mathbb{Z}\backslash\{0\}} {\pi_k(q)\over |k|}\Tr U^k\right)S_\lambda[U]&\sim &S_\lambda\{p'_k\}\ ,\nn\\
\int dU\exp\left(\sum_{k\in\mathbb{Z}\backslash\{0\}} {\pi_k(q)\over |k|}\Tr U^k\right)S_\lambda[U^{-1}]&\sim &S_\lambda\{{\bar p'}_k\}\ ,\nn\\
\int dU\exp\left(\sum_{k\in\mathbb{Z}\backslash\{0\}} {\pi_k(q)\over |k|}\Tr U^k\right)S_\lambda[U]S_\mu[U]&\sim &
S_\lambda\{p'_k\}S_\mu\{{p'}_k^{(\lambda)}\}\ ,\nn\\
\int dU\exp\left(\sum_{k\in\mathbb{Z}\backslash\{0\}} {\pi_k(q)\over |k|}\Tr U^k\right)S_\lambda[U]S_\mu[U^{-1}]&\sim &
S_\lambda\{{\bar p'}_k\}S_\mu\{{\overline{p}'}_k^{(\lambda)}\}\ ,
\ee
where $\pi_k(q) =-{(-1)^kq^{-|k|/2}\over 1-q^k}$ and bar over $p_k'$ means one is to replace $q\to q^{-1}$.
This similarly holds for particular $p_k$,
only the sum in the exponential on the l.h.s. is infinite, $i=1,\ldots, \infty$
while on the r.h.s. it is restricted by $N$:  $p'_k=\sum_{i=1}^{N} q^{k(1-i)}$ and ${p'}_k^{(\lambda)}=\sum_{i=1}^Nq^{k(\lambda_i-i+1)}$.

\bigskip

Formulas (\ref{eq:two-schurs-symmetric0})
are easy to check and generalize, as already mentioned, they are raised to a more general
identity for the Macdonald polynomials (see below).
It is also known in other branches of science under the name of Cherednik-Mehta-Macdonald (CMM) formula \cite{Che,EK,ChE}.
More difficult is to {\it understand} its interpretation and significance.
Our suggestion is to consider (\ref{eq:two-schurs-symmetric0}) as a member
of the vast family of {\it superintegrability} formulas \cite{MMsi}
for an increasing variety of eigenvalue matrix models \cite{UFN3}.
This time it is for unitary models, i.e. it generalizes (\ref{SIherm}), and does so in three directions:

\begin{itemize}

\item{}
It gives a simple bilinear average, i.e. the strict superintegrability, but significantly simpler (no linear combinations of the Schur functions are required).

\item{}
It involves a non-trivial potential, logarithm of the theta-function,
which can be considered as a deformation of the Gaussian potential for Hermitian models,
which led to a number of non-trivial developments
\cite{MMrev,CHPS,MyaM,MMP-Student,MMPSh-genBorel12}.

\item{}
As already mentioned, it has a straightforward $q,t$-deformation, which is in parallel with the $q,t$-deformation
of Hermitian matrix models \cite{MPSh,Max,China3,MMP-qt}.

\end{itemize}

Surprisingly or not, all these three directions, which were rather difficult in the Hermitian case,
become kind of simple in the unitary one.
This can have far-going implications, both conceptual and technical,
and this is why we find it reasonable to devote a special paper to this story.

\bigskip

One of such possible implications concerns the recently discussed commutative families (rays)
\cite{Ch1,Ch2,China1,China2,China3,MM-wlzz-first,MMMP-commfam,MMP-qt} in the $W_{1+\infty}$ algebra \cite{Pope,FKN2,BK,BKK,KR1,FKRN,Awata,KR2},
in the affine Yangian algebra \cite{SV,AS,Tsim,Prochazka}, and in the Ding-Iohara-Miki (DIM) algebra \cite{DI,M,A-Z}, each ray being associated with a many-body integrable system.
Providing the ordinary trigonometric Calogero-Sutherland-Ruijsenaars systems with the Schur-Jack-Macdonald eigenfunctions associated with the vertical ray,
and having, for the same ray, deep connections to the enumerative geometry
\cite{OP}
and intersection theory \cite{BSh-tau-int}: the vertical ray operators are nothing but
the completed cycles (generalized) cut-and-join operators \cite{MMN},
these commutative families acquire more sophisticated eigenfunctions for other rays, beginning with the Itzykson-Zuber
and more sophisticated unitary models.
It would be quite interesting to study their relation to the CMM formula,
especially because the latter one also has direct connections \cite{MMPchalykh} to representation theory of the DIM algebra.

\paragraph{Notation.} We use the notation $S_\lambda$ (correspondingly, $M_\lambda$) for the Schur function (the Macdonald polynomial): it is a symmetric polynomial of $N$ variables $x_i$ denoted as $S_\lambda(x_i)$ (correspondingly, $M_\lambda(x_i)$), and it is a graded polynomial of power sums $p_k=\sum_ix_i^k$ denoted as $S_\lambda\{p_k\}$ (correspondingly, $M_\lambda\{p_k\}$). Here $\lambda$ denoted the partition (or the Young diagram): $\lambda_1\ge \lambda_2\ge\ldots\ge \lambda_{l_\lambda}$, where $l_\lambda$ is the number of parts of $\lambda$. When we deal with the Schur function or the Macdonald polynomial as symmetric functions of their $N$ arguments, we {\it always} understand $l_\lambda=N$, adding if necessary some zero parts $\lambda_i=0$ at $N\ge i>l_\lambda$.

All the necessary details can be found in \cite{book:M-symmetric-functions-and-hall-polynomials,Fulton}.

We use the notation $(x;q)_\infty:=\prod_{r=0}^\infty(1-q^rx)$ for the $q$-Pochhammer symbol.

\section{The basic formulas (\ref{eq:two-schurs-symmetric0})}

\subsection{The proof}

The simplest way to derive formulas (\ref{eq:two-schurs-symmetric0}) is to use Jacobi's bialternant formula for the Schur polynomials (which is basically the first Weyl formula for characters):
\be\label{SW1}
S_\lambda(x)={\det_{i,j\le N}x_i^{N+\lambda_j-j}\over\Delta(x)} \ ,
\ee
where the Vandermonde determinant
\be
\Delta(x)=\prod_{i<j}(x_i-x_j)
\ee
so that
\be
\Delta(x^{-1})=(-1)^{N(N-1)\over 2}\prod_{i=1}^Nx_i^{1-N}\cdot\Delta(x)
\ee
Hence,
\be
\prod_{i\ne j}^N \Big(1-{z_i\over z_j} \Big)
  \ S_\lambda(z)S_\mu(z)  = (-1)^{N(N-1)\over 2}\prod_{i=1}^Nz_i^{1-N}\cdot\det_{i,j\le N}x_i^{N+\lambda_j-j}
  \cdot\det_{i,j\le N}x_i^{N+\mu_j-j}=\nn\\
  =N!(-1)^{N(N-1)\over 2}
  \sum_{\sigma,\sigma'}(-1)^{\epsilon(\sigma)+\epsilon(\sigma')}x_i^{N+1+\lambda_{\sigma(j)}-\sigma(j)+\mu_{\sigma'(j)}-
  \sigma'(j)}\ ,
\ee
where the sum runs over all permutations $\sigma$, $\sigma'$ of $N$ elements, and $\epsilon(\sigma)$ is the parity of the permutation $\sigma$. Then, using this formula and (\ref{Gth}), one immediately obtains
\be\label{a1}
& \frac{1}{N!}& \oint \prod_{i=1}^N {dz_i\over z_i}\prod_{i\ne j}^N \Big(1-{z_i\over z_j} \Big)
  \ S_\lambda(z)S_\mu(z)
  \prod_{i=1}^N\theta(z_i)
=(-1)^{N(N-1)\over 2}\sum_{\sigma}(-1)^{\epsilon(\sigma)}\prod_{i=1}^Nq^{(N+1+\lambda_i-i+\mu_{\sigma(i)}-\sigma(i))^2/2}=\nn\\
&=&(-1)^{N(N-1)\over 2}\sum_{\sigma}(-1)^{\epsilon(\sigma)}\prod_{i=1}^Nq^{(N+\psi_i(\lambda)+\psi_{\sigma(i)}(\mu))^2/2}=\nn\\
&=&(-1)^{N(N-1)\over 2}q^{N^2/2+N\sum_i(\psi_i(\lambda)+\psi_{i}(\mu))+\sum_i(\psi_i(\lambda)^2+\psi_i(\mu)^2)/2}
\sum_{\sigma}(-1)^{\epsilon(\sigma)}\prod_{i=1}^Nq^{\psi_i(\lambda)\psi_{\sigma(i)}(\mu)}\ ,
\ee
where $\psi_i(\lambda):=\lambda_i-i+1/2$.

Now using (\ref{SW1}) and the explicit formula for the quantum dimensions,
\be\label{S1}
S_\mu(q^{1-i})=(-1)^{N(N-1)\over 2}q^{-{N(N-1)(N-2)\over 6}-(N-1)\sum_j(\mu_j-j+1)}\ {\Delta(q^{\mu_i-i+1})\over\prod_{i>j} (1-q^{i-j})}\ ,
\ee
one can rewrite the r.h.s. of the first formula in (\ref{eq:two-schurs-symmetric0}) in the form
\be
S_\lambda(q^{\mu_i-i+1}) S_\mu(q^{1-i})\cdot\prod_{i>j}^N(1-q^{i-j})=
(-1)^{N(N-1)\over 2}q^{-{N(N-1)(N-2)\over 6}-(N-1)\sum_j(\mu_j-j+1)}\det_{i,j\le N}q^{(\mu_i-i+1)(N+\lambda_j-j)}=\nn\\
=(-1)^{N(N-1)\over 2}q^{-{N(N-1)(N-2)\over 6}+{N\over 4}+{1\over 2}\sum_i(\psi_i(\lambda)+\psi_i(\mu))}\det_{i,j\le N}\sum_{\sigma}(-1)^{\epsilon(\sigma)}\prod_{i=1}^Nq^{\psi_i(\lambda)\psi_{\sigma(i)}(\mu)}\ .
\ee
Comparing this expression with (\ref{a1}), one immediately obtains the first formula in (\ref{eq:two-schurs-symmetric0}).

The second formula in (\ref{eq:two-schurs-symmetric0}) is proved absolutely the same way.

\subsection{Simple examples and expansions}

While very concise in their form (\ref{eq:two-schurs-symmetric0}),
the relations are in fact rather cumbersome in details.
In order to get a flavour of formulas (\ref{eq:two-schurs-symmetric0}), we consider some simple examples and expansions.

\begin{itemize}

\item{\bf $q$-expansion}

The simplest way to validate (\ref{eq:two-schurs-symmetric0}) is to look at the expansion in powers of $q$.
Then only a finite number of terms in the sum over $k$ contribute at given $N$, $\lambda$ and $\mu$.

At $q=0$, the theta-function (\ref{thetaf0}) becomes unity, and we get the ordinary orthogonality relation
\eqref{orthoU} for the unitary model,
with $|\lambda|!\cdot \delta_{\lambda,\mu}$ at the r.h.s.
(see, for instance \cite[p.370]{book:M-symmetric-functions-and-hall-polynomials}).

In the first order of the $q$-expansion, $\prod_i \theta(z_i) = 1 + q^{1/2}\left( S_{[1]}(z) + S_{[1]}(z^{-1})\right) + O(q)$,
one can apply the Pieri rule for the product $S_\lambda S_{[1]} = \mathop{\sum}_{\lambda^{'} = \lambda + \Box} S_{\lambda^{'}}$
so  that the answer is
$q^{1/2}\left( \delta_{\lambda,\mu + \Box} + \delta_{\lambda + \Box,\mu}\right)$.
This answer is already non-trivial to extract from the r.h.s. of \eqref{eq:two-schurs-symmetric0}:
the factor $S_\lambda(q^{-\mu+\dots})$ contains negative powers of $q$, while $S_\mu(q^{\dots})$
contains positive powers of $q$, so several summands combine to produce the result.

\item{\bf Particular $N$}

 At $N=1$, only the Schur functions with single row Young diagrams do not vanish, therefore, $\lambda$ and $\mu$ are just natural numbers, and $S_\lambda(z)=z^\lambda$.
Then, the average at the l.h.s. of (\ref{eq:two-schurs-symmetric0} is just
\be
\sum_{k=-\infty}^\infty  q^{k^2/2} \oint \frac{dz}{z} z^{\lambda-\mu} z^k
=  \sum_{k=-\infty}^\infty q^{k^2/2}\delta_{\lambda-\mu+ k,0}=  q^{(\lambda-\mu)^2/2}\ ,
\ee
 and the r.h.s. is also
\be
q^{c_{\lambda,\mu}}q^{(1-\lambda)\mu}q^\lambda = q^{-(\lambda+\mu)/2+\lambda(\lambda-1)/2+\mu(\mu-1)/2+(1-\lambda)\mu+\lambda }
= q^{(\lambda-\mu)^2/2}\ .
\ee

\item{\bf Heat equation for $\theta$}

One can apply the formula
$q\frac{\p}{\p q} \prod_i \theta(z_i) =\frac{1}{2}  \sum_i\left( z_i\frac{\p}{\p z_i}\right)^2 \prod_i \theta(z_i)$
and integrate by parts.

In the simplest case of $N=1$,
\be
\sum_k \frac{k^2}{2}q^{k^2/2}\delta_{\lambda-\mu+ k,0}= \frac{(\lambda-\mu)^2}{2}q^{(\lambda-\mu)^2/2}
=\frac{1}{2}\sum_k q^{k^2/2}(\lambda-\mu)^2 \delta_{\lambda-\mu+k,0}\ .
\ee

The typical examples of Schur functions at special points are (the Young diagrams here have to be padded with zeroes to have exactly $N$ lines)
\be
\lambda=[3,1,1], \mu=[5]: &
S_{\lambda}(q^{-\mu_i+i-1}) =S_{[3,1,1]}(q^{-5\delta_{i,1}+i-1}) = \frac{(q^2 + 1)(q^{12} + q^{11} - q^9 + q^7 + q^4 - q^2 + 1)}{q^7}\ ,
\nn \\
\lambda=[1,1], \mu=[3,1]: &
S_{\lambda}(q^{-\mu_i+i-1}) =S_{[1,1]}(q^{-3\delta_{i,1}-\delta_{i,2}+i-1})
= \frac{q^{10}-q^8+2q^6+q^2-1}{2q^6}\ , \nn \\
\ldots
\label{Sspec}
\ee
Also the two sides of heat equation, i.e. the $\log q$ derivative of the r.h.s.
and the average of Laplace action at the l.h.s. look like (for single-line $\lambda$ and empty $\mu=\emptyset$):

{\tiny
\be
\!\!\!\!\!\!\!\!\!\!\!\!\!\!\!\!\!\!\!\!\!\!\!\!\!\!\!\!\!\!\!
\begin{array}{c||c|c|c|c}
\lambda & N=1 & N=2 & N=3 & \ldots \\
\hline\hline
&&&& \\
1 & q^{1/2} &-q^{1/2}\Big(5q^2-1\Big)&-q^{1/2}(q-1)^2\Big(13q^4+15q^3+8q^2-q-1\Big)& \\
&&&& \\
2 & 4q^2 &-2q^2\Big(5q^3-2\Big)&-2q^2(q-1)^2\Big(10q^6+11q^5+12q^4+6q^3-q-2\Big)&\\
&&&& \\
3 & 9q^{9/2} &-q^{9/2}\Big(17q^4-9\Big)&-q^{9/2}(1-q)^2\Big(29q^8+31q^7+33q^6+35q^5+16q^4-3q^3-5q^2-7q-9\Big)& \\
&&&& \\
4 & 16 q^8 &-2q^8\Big(13q^5-8\Big)&-2q^8(1-q)^2\Big(20q^{10}+21q^9+22q^8+23q^7+24q^6+10q^5-4q^4-5q^3-6q^2-7q-8\Big)& \\
&&&& \\
5 & 25 q^{25/2} &-q^{25/2}\Big(37q^6-25\Big) &-q^{25/2}(1-q)^2\Big(53q^{12}+55q^{11}+57q^{10}+59q^9+61q^8+63q^7+
+24q^6-15q^5-17q^4-19q^3-21q^2-23q-25\Big) & \\
&&&& \\
\ldots &\ldots &\ldots &\ldots &
\end{array}
\nn
\ee
}

\end{itemize}

These are just a handful of small illustrations demonstrating that (\ref{eq:two-schurs-symmetric0}), though being easy to prove,
is highly non-trivial.
One can only admire the coincidence of structures like this at two sides of the heat equation,
or appearance of structures like (\ref{Sspec}) from the ``innocent" integrals over $z$ at the l.h.s.
of (\ref{eq:two-schurs-symmetric0}).
This is a nice illustration of {\bf the power of superintegrability}, and stimulates search
for {\bf a hidden symmetry} which underlies it: for this particular model,
and for unitary matrix models in general.

\subsection{Implications}

This seeming complexity has two rationalizations/interpretations.

{\bf First}, it demonstrates that averages in the unitary matrix model can be rather non-trivial.
This is despite the fact that the unitary matrix model possesses superintegrability property \cite{MMsi}
like the Gaussian Hermitian one (which is quite interesting by itself,
since the integral is no longer Gaussian).
But this is best seen for an average of a {\it single} Schur, which is indeed
much simpler.
In fact, pair correlators are somewhat sophisticated even in the Hermitian model
\cite{MMsi},
but not as much as \eqref{Sspec}.
Instead, here we deal with two ``pure'' Schur functions, without a need to
modify one of them as in \cite{MMsi}.

{\bf Second}, the averages (\ref{eq:two-schurs-symmetric0}) have a direct physical meaning:
they reproduce certain correlators in Chern-Simons theory, that is,
the Hopf HOMFLY-PT polynomials colored with two arbitrary representations,
and these are expected to be complicated in a non-trivial way:
they should be strong enough knot/link invariants: the hypothesis
that a set of all colored HOMFLY-PT polynomials forms the complete knot invariant
is not currently disproved.

In what follows, we discuss in more detail the superintegrability properties
of unitary model and two kinds/directions of its deformation:
(i) to other knots, actually to the torus ones, the so called TBEM model \cite{BEMT}
(its further generalization beyond the torus knots faces some strange problems \cite{AMMM}),
and (ii) a deformation from Schur to Jack and Macdonald functions.
Surprisingly or not, it turns out to be rather difficult to reconcile these two deformations (i) and (ii).
While (ii) is nice for the Hopf link, moreover, is promisingly formalized and promoted
to the simple CMM formula in terms of residues, it does not work quite the same way
for other links and knots, even for the torus ones.

In sec.\ref{sec:unimat}, we discuss superintegrable formula
\eqref{eq:two-schurs-symmetric0} from the perspective of unitary matrix models.
In sec.\ref{sec:qt-def}, we provide, from very general considerations, a
\textit{reasonable} form for the $t$-deformation of \eqref{eq:two-schurs-symmetric0}.
In sec.\ref{sec:cmm-formulas}, we explain that this reasonable guess is, indeed,
correct, the resulting formulas known under the name of Cherednik-Macdonald-Mehta
identities are intricately connected with representation theory of double affine Hecke algebras and,
as such, naturally give \textit{hyperpolynomials} for the Hopf link
(in two possible mutual orientations of components).
In sec.\ref{sec:bemt-model}, we highlight the problems in generalizing the TBEM model to the CMM like formulas:
an expected generalization of the CMM formulas to arbitrary $(m,n)$ besides $(2,2)$,
does not reproduce even the unknot hyperpolynomials,
not to mention more complicated torus knots/links.

\section{Unitary matrix models} \label{sec:unimat}

The unitary one-matrix model is given by the invariant integral over the $N\times N$ unitary matrix with the Haar measure $dU$ invariant under multiplication by a unitary matrix:
\be
Z_N\Big(\{p_k\}_{k\in\mathbb{Z}\backslash\{0\}}\Big)=\int dU\exp\left(\sum_{k\in\mathbb{Z}\backslash\{0\}} {p_k\over |k|}\Tr U^k\right)\ .
\ee
This integral considered as a power sum in $p_k$ can be presented as an infinite sum \cite{MMZ}
\be\label{main}
Z_N\Big(\{p_k\}_{k\in\mathbb{Z}\backslash\{0\}}\Big)=\sum_{\lambda:\ l_\lambda\le N}S_\lambda\{p_k\}S_\lambda\{p_{-k}\}\ .
\ee
This partition function provides us with a generating function of averages in the trivial background. 

In particular, one can get the average (we define the normalization of the measure in such a way that $\langle 1\rangle=1$)
\be
Z_N\Big(\{p_k\}_{k\in\mathbb{Z}\backslash\{0\}}\Big)=\langle S_\lambda[U]S_\mu[U^{-1}]\rangle_{0}:=\int dUS_\lambda[U]S_\mu[U^{-1}]=\delta_{\lambda\mu}\ ,
\ee
where we use the notation $S_\lambda[U]:=S_\lambda\{\Tr U^k\}$.

The generating function of averages in a non-trivial background is given by the shift $p_k\to p_k+P_k$ in (\ref{main}) with some non-zero fixed values of $P_k$, while $p_k$ continue to play the role of generating parameters. However, the averages associated with non-zero $P_k$ are generally given by infinite sums \cite{MMZ}:
\be
\langle S_\lambda[U]S_\mu[U^{-1}]\rangle_{P_k}:=\int dU\exp\left(\sum_{k\in\mathbb{Z}\backslash\{0\}} {P_k\over |k|}\Tr U^k\right)S_\lambda[U]S_\mu[U^{-1}]=\sum_{\nu:\ l_\nu\le N}S_{\nu/\lambda}\{P_k\}S_{\nu/\mu}\{P_{-k}\}\ ,
\ee
where $S_{\nu/\lambda}$ denotes the skew Schur function \cite{book:M-symmetric-functions-and-hall-polynomials}.
Still, there is (at least one) distinguished case when this sum also becomes a finite polynomial: when the background field is equal to
\be
P_k=\pi_k(q)=-{(-1)^kq^{-|k|/2}\over 1-q^k},\ \ \ \ \ k\in\mathbb{Z}\ ,
\ee
we come to formulas (\ref{nb}).

Indeed, one may note that integrating over the angular variables
\cite{Mehta} gives rise to
\be\label{ei}
Z_N\Big(\{p_k\}_{k\in\mathbb{Z}\backslash\{0\}}\Big)={1\over N!}\prod_{i=1}^N\oint {dz_i\over z_i}\Delta(z)\Delta(z^{-1})
\exp\left(\sum_{i=1}^N\sum_{k\in\mathbb{Z}\backslash\{0\}} {p_k\over |k|}z_i^k\right)\ ,
\ee
where $\Delta(z)=\prod_{i<j}(z_i-z_j)$ is the Vandermonde determinant, while the averages are given by
\be
\langle S_\lambda(z_i)S_\mu(z_i^{\pm 1})\rangle_{P_k}={1\over N!}\prod_{i=1}^N\oint {dz_i\over z_i}
\underbrace{\Delta(z)\Delta(z^{-1})}_{\prod_{i\ne j}^N\left(1-{z_i\over z_j}\right)}
\exp\left(\sum_{i=1}^N\sum_{k\in\mathbb{Z}\backslash\{0\}} {P_k\over |k|}z_i^k\right)S_\lambda(z_i)S_\mu(z_i^{\pm 1})\ .
\ee
Now let us choose the background to be $P_k=\pi_k$,
\be\label{pi}
\pi_k(q)=-{(-1)^kq^{-|k|/2}\over 1-q^k},\ \ \ \ \ k\in\mathbb{Z}\ .
\ee
Then, using formula (\ref{thetaf0}), the Jacobi triple product identity, which is valid for $|q|<1$, $z \neq 0$
\begin{align}\label{theta}
  \theta(z) = \prod_{m=1}\left(1 - q^m\right)
  \left(1 + \frac{z}{q^{1/2}} q^m \right)
  \left(1 + \frac{1}{z q^{1/2}} q^m \right)
  = \left(q ;q\right)_\infty \left(-\frac{z}{q^{1/2}} ;q\right)_\infty
  \left(-\frac{1}{z q^{1/2}} ;q\right)_\infty\ ,
\end{align}
and the identity
\be\label{Pe}
(z;q)_\infty=\exp\left[-\sum_m{z^m\over (1-q^m)m}\right]\ ,
\ee
one obtains that formulas (\ref{eq:two-schurs-symmetric0}) give rise to 
\be
\langle S_\lambda(z_i)S_\mu(z_i^{\pm 1})\rangle_{\pi_k}=q^{c^\pm_{\lambda,\mu}}
    \cdot S_\lambda(q^{\pm(\mu_i-i+1)}) S_\mu(q^{\pm(1-i)})\cdot\prod_{i>j}^N(1-q^{i-j})\ ,
\ee
and can be associated with the strict superintregrability of the unitary matrix model in the background $\pi_k$.

\section{$q,t$-deformation} \label{sec:qt-def}

In order to construct a $q,t$-deformation of the unitary matrix model, we have to deform the eigenvalue integral (\ref{ei}), since such a deformation typically does not admit a matrix integral representation anymore, see an example of the $q,t$-deformed Hermitian model in
\cite{MPSh,Max,MMell}. At the same time, from this example one can expect that the superintegrability is preserved with replacing of the Schur functions by the Macdonald polynomials. Fortunately, in the case of unitary matrix model, we know even how the strict superintegrability has to look, since we know a proper $q,t$-deformation of the r.h.s. of formulas (\ref{eq:two-schurs-symmetric0}): in the non-deformed case, they are given by the HOMFLY polynomial of the Hopf link, hence, we expect in the deformed case the hyperpolynomial of the Hopf link $P_{\lambda,\mu}^{Hopf}(q,t,N)$ \cite{IK}, which is also known \cite{Chek}, say, from theory of DAHA
\cite{Ch-daha-and-mac}
(see also \cite[Eq.(39)]{AKMM1}). In other words, the superintegrability in the $q,t$-deformed case looks as
\be
\langle M_\lambda(z_i)M_\mu(z_i^{\pm 1})\rangle\sim P_{\lambda,\mu}^{Hopf}(q,t,N)\ .
\ee
The explicit expression for this hyperpolynomial is
\be
P_{\lambda,\mu}^{Hopf}(q,t,N)\sim f_\mu f_\lambda
    M_\mu(q^{-\lambda_i}t^i) M_\lambda(t^i)\ ,
\ee
where $\nu_\lambda:=2\sum_i(i-1)\lambda_i$, $\nu'_\lambda:=\nu_{\lambda^\vee}$, and the quantity $f_\lambda=q^{\nu'_\lambda/2}t^{-\nu_\lambda/2}$ is the standard Taki's framing factor \cite{Taki}. Note that $M_\lambda(t^{i-1})$ is the Macdonald dimension \cite{AKMM1} (which generalizes (\ref{S1})):
\be
M_\lambda(t^{i})=t^{\sum_ii\lambda_i}\cdot
\prod_{i<j}{(q^{\lambda_i-\lambda_j}t^{j-i};q)_\infty\over (q^{\lambda_i-\lambda_j}t^{j-i+1};q)_\infty}
{(t^{j-i+1};q)_\infty\over (t^{j-i};q)_\infty}\ \stackrel{t=q^k}{=}\ t^{\sum_ii\lambda_i}\cdot
\prod_{r=0}^{k-1}\prod_{i<j}{1-q^{r+\lambda_i-\lambda_j}t^{j-i}\over 1-q^rt^{j-i}}\ ,
\ee
and $M_\mu(q^{-\lambda_i}t^{i-1})$ can be rewritten in terms of power sums extending (\ref{nb}): $M_\mu\{{p'}_k^{(\lambda)}\}$, where \cite[formula (30)]{AKMM1}
\be
{p'}_k^{(\lambda)}:=\sum_{i=1}^nq^{-k\lambda_i}t^{k(i-1)}
={1-t^{kN}\over 1-t^k}+\sum_{i=1}t^{k(i-1)}\Big(q^{-k\lambda_i}-1\Big)\ .
\ee

In fact, we know even more: from the example of the Hermitian model
\cite{MPSh,LPSZ}, it is known how to deform the product of two Vandermonde determinants:
\be\label{V1}
\Delta(z)\Delta(z^{-1})=\prod_{i\ne j}^N\left(1-{z_i\over z_j}\right)\longrightarrow\prod_{i\ne j}^N{(z_i/z_j;q)_\infty\over(tz_i/z_j;q)_\infty}\ ,
\ee
where $(x;q)_\infty:=\prod_{r=0}^\infty(1-q^rx)$ is the $q$-Pochhammer symbol. In the case of $t=q^k$ with an integer $k$, it becomes
\be\label{V2}
\Delta(z)\Delta(z^{-1})=\prod_{i\ne j}^N\left(1-{z_i\over z_j}\right)\longrightarrow\prod_{i\ne j}^N\prod_{r=0}^{k-1}\left(1-q^r{z_i\over z_j}\right)\ .
\ee

It remains only to fix the potential in the $q,t$-deformed model. In the case of the Hermitian model, it is obtained by the $q$-deformation of the Gaussian exponential to the $q$-exponential \cite{MPSh}. What is important, the deformation of the potential does not contain $t$ but only $q$. And, in the unitary model that we discuss here, the potential (\ref{pi}) is already $q$-dependent, and changing the Hopf HOMFLY-PT polynomial for the hyperpolynomial gives rise to a $t$-deformation only. This means that one has to change
\textit{only the Vandermonde factors keeping the potential intact}. Thus, one expects the superintegrability in the $q,t$-deformed unitary model in the form
\be\label{qtu}
\boxed{
{1\over N!}\prod_{i=1}^N\oint {dz_i\over z_i}\prod_{i\ne j}^N{(z_i/z_j;q)_\infty\over(tz_i/z_j;q)_\infty}
\exp\left(\sum_{i=1}^N\sum_{k\in\mathbb{Z}\backslash\{0\}} {\pi_k(q)\over |k|}z_i^k\right)M_\lambda(z_i)M_\mu(z_i^{-1})\sim P_{\lambda,\mu}^{Hopf}(q^{-1},t^{-1},N)}\ ,
\ee
where the r.h.s. is the hyperpolynomial of the mirror reflected Hopf link.

In the next section, we check this kind of formulas and demonstrate that, for the average $\langle M_\lambda(z_i)M_\mu(z_i)\rangle$, instead of using the hyperpolynomial of Hopf link, it is better still to use the hyperpolynomial of the mirror reflected Hopf link, but to make a reflection $q\to q^{-1}$ in the potential: $\pi_k(q)\to \pi_k(q^{-1})=-{(-1)^kq^{|k|/2}\over 1-q^{-k}}$ (the reflection $q\to q^{-1}$ makes a mirror reflection of the Hopf link at $t=q$, but at $t\ne q$ does not).

\section{CMM formulas} \label{sec:cmm-formulas}

In fact, it turns out that formulas of the (\ref{qtu}) kind are known, they were found by I. Cherednik \cite{Che} as an extension of a formula due to I.G. Macdonald and Mehta \cite{MM-conj} and are since called CMM formulas. Another approach to them can be found in
\cite{Cha,ChaF}. These formulas have basically two forms at $t=q^k$ with an integer $k$:
\be\label{CMM}
 & \frac{1}{N!}& \oint \prod_i{dz_i\over z_i}\prod_{i\ne j}^N\prod_{r=0}^{k-1}\left(1-q^r{z_i\over z_j}\right)M_\lambda(z)M_\mu(z^{-1})\prod_i\theta(z_i)=\nn\\
&  =&
      q^{|\lambda| + |\mu|\over 2}f_\mu f_\lambda
    M_\mu(q^{-\lambda_i}t^{i-1}) M_\lambda(t^{i-1})\prod_{i\ne j}^N\prod_{r=0}^{k-1}\left(1-q^rt^{i-j}\right)\\
 & \frac{1}{N!}& \oint \prod_i{dz_i\over z_i}\prod_{i> j}^N\prod_{r=0}^{k-1}\Big(1-q^r{z_i\over z_j}\Big)M_\lambda(z)M_\mu(z)\prod_i\theta^{(-)}(z_i)=\nn\\
 & =&(-1)^{N(N-1)\over 2}t^{{N(N-1)\over 12}(2kN-k+3)}t^{(N-1)(|\mu|+|\lambda|)}
      q^{|\lambda| + |\mu|\over 2}f_\mu f_\lambda
    M_\mu(q^{-\lambda_i}t^{i-1}) M_\lambda(t^{i-1})\prod_{i> j}^N\prod_{r=0}^{k-1}\Big(1-q^rt^{i-j}\Big)\ ,\nn
\ee
where
\be
\theta^{(-)}(z):=\sum_nq^{-n^2/2}z^{n}\ .
\ee
The symmetricity of these formulas in $\lambda$ and $\mu$, which is definitely associated with symmetry under permutation of the two components of the Hopf link, follows from the two facts:
\begin{itemize}
\item The invariance of the Macdonald polynomials under the transform $(q,t)\to (q^{-1},t^{-1})$ \cite{book:M-symmetric-functions-and-hall-polynomials}.
\item The duality of the Macdonald polynomials \cite{book:M-symmetric-functions-and-hall-polynomials}:
\be
{M_\lambda(q^{\mu_i}t^{-i})\over M_\lambda(t^{-i})}={M_\mu(q^{\lambda_i}t^{-i})\over M_\mu(t^{-i})}\ .
\label{duaid}
\ee
\end{itemize}

Formulas (\ref{CMM}) can be definitely rewritten in terms of Gaussian integrals using (\ref{Gth}), and can be also rewritten in terms of the background $\pi_k(q)$ (\ref{pi}) reproducing (\ref{qtu}):
\be
{1\over N!}\prod_{i=1}^N\oint {dz_i\over z_i}\prod_{i\ne j}^N{(z_i/z_j;q)_\infty\over(tz_i/z_j;q)_\infty}
\exp\left(\sum_{i=1}^N\sum_{k\in\mathbb{Z}\backslash\{0\}} {\pi_k(q)\over |k|}z_i^k\right)M_\lambda(z_i)M_\mu(z_i^{-1})\sim P_{\lambda,\mu}^{Hopf}(q^{-1},t^{-1},N)\ ,\nn\\
{1\over N!}\prod_{i=1}^N\oint {dz_i\over z_i}\prod_{i\ne j}^N{(z_i/z_j;q)_\infty\over(tz_i/z_j;q)_\infty}
\exp\left(\sum_{i=1}^N\sum_{k\in\mathbb{Z}\backslash\{0\}} {\pi_k(q^{-1})\over |k|}z_i^k\right)M_\lambda(z_i)M_\mu(z_i)\sim P_{\lambda,\mu}^{Hopf}(q^{-1},t^{-1},N)\ .
\ee

Note that, in contrast to formulas (\ref{eq:two-schurs-symmetric0}), which are proved directly, available proofs of the CMM formulas are far more involved, since simple determinant representations do not exist for the Macdonald polynomials. However, a counterpart of the Weyl formula does exist for them too, and is related to the representations of the Macdonald polynomials via the (generalized, multivariate) Baker-Akhiezer functions \cite{Cha,ChaF}. This approach allows one to construct a relatively immediate proof of the CMM formulas \cite{ChE}. Other proofs rely heavily on using the double affine Hecke algebras \cite{Che} or representation theory of quantum groups \cite{EK}.

\section{Problems with $q,t$-deformation of TBEM model} \label{sec:bemt-model}

Formulas (\ref{uniHopf}) and (\ref{eq:two-schurs-symmetric0})  have an interesting generalization \cite{BEMT},
which describes arbitrary unnormalized HOMFLY polynomials for $L$-component torus links $(Lm,Ln)$
\be\label{TBEM}
H_{\lambda_1,\ldots, \lambda_L}^{(Lm,Ln)}(A=q^N,q) \sim &
\frac{1}{N!} \oint \prod_{i=1}^N {dz_i\over z_i} z_i^{\kappa_{m,n}} \cdot \Delta_m(z) \Delta_n(z)
  \prod_{a=1}^L S_{\lambda_a}(z_i^{m n})
  \prod_{i=1}^N\theta_{2m n}(z_i)\ , &
  \\ \nn
  \theta_{a} (z) := \sum_{k=-\infty}^\infty q^{k^2/a}z^{k}\ ,
  \ \ \ \ &
  \kappa_{m,n}: = -{(m+n)(N-1)\over 2}\ ,\ \ \ \ \ \ \Delta_m(z) := \prod_{i<j}^N \left(z_i^m-z_j^m\right)\ .
\ee
Its deformation to arbitrary knots, even to the twist family, faces some mysterious obstacles \cite{AMMM}.
For all cases besides the full-twist family $(m,n)=(1,1)$
(that includes the Hopf link), it is no longer a unitary model,
because of deformation of the Vandermonde determinants $\Delta(z)\longrightarrow\Delta_m(z)$.

Let us explain how to derive formula (\ref{TBEM}) in the simplest case of a torus knot $(m,n)$, i.e. at $L=1$. We use the trick by \cite{BEMT}. The derivation consists of two steps.
\begin{itemize}
\item First, we start with the integral (\ref{TBEM}) with $m=1$ and $n=\alpha$ with rational $\alpha$:
\be
I_\lambda(\alpha):=\frac{1}{N!} \oint \prod_{i=1}^N {dz_i\over z_i} z_i^{\kappa_{1,\alpha}} \cdot \Delta(z) \Delta_\alpha(z)
  S_{\lambda}(z_i^{\alpha})
  \prod_{i=1}^N\theta_{2\alpha}(z_i)\ .
\ee
 Then, using (\ref{SW1}), one obtains
\be\label{qdim1}
I_\lambda(\alpha)=\frac{1}{N!} \oint \prod_{i=1}^N {dz_i\over z_i} z_i^{\kappa_{1,\alpha}} \cdot \det_{i,j\le N}z_i^{\alpha(N+\lambda_j-j)}
\det_{i,j\le N}z_i^{(j-1)}
  \prod_{i=1}^N\theta_{2\alpha}(z_i)\sim\nn\\
  \sim q^{\alpha\kappa_2(\lambda)}\sum_\sigma(-1)^{\epsilon(\sigma)}\prod_iq^{(\sigma(i)-1)(N+\lambda_i-i)}
  \sim q^{\alpha\kappa_2(\lambda)}\Delta(q^i)S_\lambda(q^{i-1})\ ,
\ee
where we used (\ref{Gth}) assuming odd integer $\alpha$ and then continued it to an arbitrary rational number (one can also make this calculation for an arbitrary rational $\alpha$ in terms of the Gaussian integral (\ref{Gth}) \cite{BEMT}). Thus,
\be\label{qdim}
{I_\lambda(\alpha)\over I_\emptyset(\alpha)}\sim q^{\alpha\kappa_2(\lambda)}S_\lambda(q^{i-1})\sim
q^{\alpha\kappa_2(\lambda)}\ \hbox{Unknot}_\lambda\ ,
\ee
which is the quantum dimension with a framing factor.
\item
Second, we make the change of variables $z_i\to z_i^{1\over m}$ in (\ref{TBEM}) and to choose $\alpha=n/m$. This gives rise to the integral
\be\label{qmI}
I_\lambda(m,n):=\frac{1}{N!} \oint \prod_{i=1}^N {dz_i\over z_i} z_i^{{1\over m}\kappa_{m,n}} \cdot \Delta(z) \Delta_\alpha(z)
  S_{\lambda}(z_i^{\alpha m})
  \prod_{i=1}^N\theta_{2\alpha}(z_i)\ ,
\ee
since moments of $\theta_{2\alpha m^2}(z_i^{1\over m})$ and those of $\theta_{2\alpha}(z_i)$ coincide. In order to estimate this integral, we expand
\be
S_{\lambda}(z_i^{\alpha m})=\sum_{\mu\vdash m|\lambda|}c^\mu_\lambda S_\mu(z_i^\alpha)\ ,
\ee
where $c^\mu_\lambda$ are the Adams coefficients, and integral (\ref{qmI}) reduces to the sum of integrals (\ref{qdim1}).
Using now (\ref{qdim}), one immediately obtains
\be
{I_\lambda(m,n)\over I_\emptyset(m,n)}\sim
\sum_{\mu\vdash m|\lambda|}c^\mu_\lambda q^{{n\over m}\kappa_2(\mu)}S_\mu(q^{i-1})\ .
\ee
The r.h.s. of this formula is nothing but the notorious Rosso-Jones formula for the HOMFLY-PT polynomial of the torus knot $(m,n)$ \cite{RJ,China}.
\end{itemize}

One may ask if there an extension of this construction to the hyperpolynomials of the torus knots similarly to the case of the Hopf link. A natural candidate for the $t$-deformation of the first step would be the formula ($t=q^k$)
\be\label{1}
I_\lambda^{(t)}(\alpha):=\frac{1}{N!} \oint \prod_{i=1}^N {dz_i\over z_i} {\cal D}_\alpha^{(q,k)}(z)
  M_{\lambda}(z_i^{\alpha})
  \prod_{i=1}^N\theta_{2\alpha}(z_i)\ ,
\ee
where ${\cal D}_\alpha^{(q,k)}(z)$ is a deformation of the products:
\be\label{p1}
{\cal D}_1^{(q,k)}(z)=\prod_{i\ne j}\prod_{r=0}^{k-1}\left(1-q^r{z_i\over z_j}\right)
\ee
at $\alpha=1$, and
\be\label{p2}
{\cal D}_\alpha^{(q,1)}(z)=\prod_{i< j}\left(z_i- z_j\right)\prod_{i<j}\left(z_i^\alpha- z_j^\alpha\right)
\ee
at $k=1$. The key problem is that it is not easy to deform these two products, which are symmetric functions of $z_i$ in such a way that it will be still a symmetric function, and $\alpha\ne 1$, $k\ne 1$. One can check that (\ref{1}) with non-symmetric combinations like
\be
{\cal D}_\alpha^{(q,k)}(z)= \left(\prod_iz_i^{\kappa_{\ast}}\right) \cdot \prod_{i>j}\prod_{r=0}^{k-1}(z_i-q^rz_j) \prod_{i<j}\prod_{r=0}^{k-1}(z_i^\alpha-q^rz_j^\alpha)
 \ee
 or
 \be
 {\cal D}_\alpha^{(q,k)}(z)= \left(\prod_iz_i^{\kappa_{\ast}}\right) \cdot \prod_{i<j}\prod_{r=0}^{k-1}(z_i-q^rz_j) \prod_{i>j}\prod_{r=0}^{k-1}(z_i^\alpha-q^rz_j^\alpha)
 \ee
with some $\kappa_\ast$ equal to $-2k$ at $\alpha=1$ are expectedly not factorized at any reasonable $\kappa_\ast$.

Thus, at the moment the TBEM generalization of CMM and matrix model representation of torus hyperpolynomials
beyond examples of colored hyperpolynomials for the Hopf link remains a puzzle.

\section{Conclusion}

To conclude, we found a place for the CMM formulas in the emerging general theory of superintegrability.
They provide a $q,t$-deformation of unitary model and thus describe colored Hopf hyperpolynomials.
A number of interesting unsolved problems,
which look more prominent in light of this understanding, are listed in the text,
they open a big research direction for the future.
This also calls for a re-examination of the origins of CMM formula, which is a separate interesting story,
and will be presented elsewhere \cite{MMPchalykh}.

\section*{Acknowledgements}

This work was funded through the state assignment of the NRC Kurchatov Institute and partly supported by the grants of the Foundation for the Advancement of Theoretical Physics and Mathematics ``BASIS" and by joint grant 21-51-46010-ST-a.

\end{document}